\pdfoutput=1
\documentclass[aps,prd,floats,twocolumn,epsf,nofootinbib,showpacs,superscriptaddress]{revtex4-1}
\usepackage{graphicx}
\usepackage{bm}
\usepackage{times}
\usepackage{cancel}
\usepackage{comment}
\usepackage{slashed}
\usepackage{color}
\usepackage{slashed}
\usepackage{lipsum}
\usepackage{subfigure}
\usepackage{multirow}
\usepackage{amsmath}
\usepackage{array} 
\usepackage{varwidth} 
\usepackage[dvipsnames]{xcolor}

\newcommand{\be}{\begin{equation}}
\newcommand{\ee}{\end{equation}}
\newcommand{\bea}{\begin{eqnarray}}
\newcommand{\eea}{\end{eqnarray}}

\begin{document}

\title{
Cosmic Ray Antihelium from a Strongly Coupled Dark Sector
}
\author{Martin Wolfgang Winkler}
\email{martin.winkler@su.se, ORCID: orcid.org/0000-0002-4436-0820}
\affiliation{Department of Physics, The University of Texas at Austin, Austin, 78712 TX, USA}
\affiliation{The Oskar Klein Centre, Department of Physics, Stockholm University, AlbaNova, SE-10691 Stockholm, Sweden}
\author{Pedro De La Torre Luque}
\email{pedro.delatorreluque@fysik.su.se, ORCID: 0000-0002-4150-2539}
\affiliation{The Oskar Klein Centre, Department of Physics, Stockholm University, AlbaNova, SE-10691 Stockholm, Sweden}
\author{Tim Linden}
\email{linden@fysik.su.se, ORCID: orcid.org/0000-0001-9888-0971}
\affiliation{The Oskar Klein Centre, Department of Physics, Stockholm University, AlbaNova, SE-10691 Stockholm, Sweden}

\begin{abstract}
Standard Model extensions with a strongly coupled dark sector can induce high-multiplicity states of soft quarks. Such final states trigger extremely efficient antinucleus formation. We show that dark matter annihilation or decay into a strongly coupled sector can dramatically enhance the cosmic-ray antinuclei flux -- by six orders of magnitude in the case of ${^4\overline{\text{He}}}$. In this work, we argue that the tentative ${^3\overline{\text{He}}}$ and ${^4\overline{\text{He}}}$ events reported by the AMS-02 collaboration could be the first sign of a strongly coupled dark sector observed in nature.
\end{abstract}

\maketitle

\section{Introduction}

Cosmic-ray (CR) antinuclei are among the most promising targets in the indirect search for particle dark matter (DM). While the formation of antinuclei by DM annihilation or decay is strongly suppressed compared to {\it e.g.}\ gamma rays, the astrophysical antinuclei backgrounds -- which arise from interactions of cosmic ray protons and helium with the interstellar gas -- are extremely low. Therefore, the unambiguous discovery of even a single cosmic-ray antinucleus could provide smoking-gun evidence for particle DM~\cite{Donato:1999gy,Baer:2005tw}.

While antideuterons have long been the prime target for cosmic-ray antinuclei searches~\cite{Battiston:2008zza,Fuke:2005it,Aramaki:2015laa}, the antihelium channel has recently gained attention because the Alpha Magnetic Spectrometer (AMS-02) has tentatively detected at least 8~$\overline{\text{He}}$ events~\cite{Ting:2016,Choutko:2018,vonDoetinchem:2020vbj}. 
While 6 of the candidates are more likely $^3\overline{\text{He}}$, 2 lie in the mass range for $^4\overline{\text{He}}$ -- although a single isotopic origin is not excluded due to the mass resolution of AMS-02.\footnote{Recent results include 9 antihelium candidate events which are evenly distributed between $^3\overline{\text{He}}$ and $^4\overline{\text{He}}$~\cite{Zuccon:2022}.} This observation was unexpected because (1) the astrophysical antihelium flux is expected to fall at least an order of magnitude below the AMS-02 sensitivity~\cite{Chardonnet:1997dv,Duperray:2005si,Herms:2016vop,Blum:2017qnn,Korsmeier:2017xzj,Poulin:2018wzu,Shukla:2020bql}, and (2) DM annihilation seemed to predict an antihelium flux far below the observed rate once indirect detection constraints are taken into account ~\cite{Cirelli:2014qia,Carlson:2014ssa,Coogan:2017pwt,Kachelriess:2020uoh}, unless the dark sector is finely tuned~\cite{Heeck:2019ego}.$\;\;$
\begin{figure}[ht!]
\centering
\includegraphics[trim={1mm 1mm 1mm 1mm},clip,width=.285\textwidth]{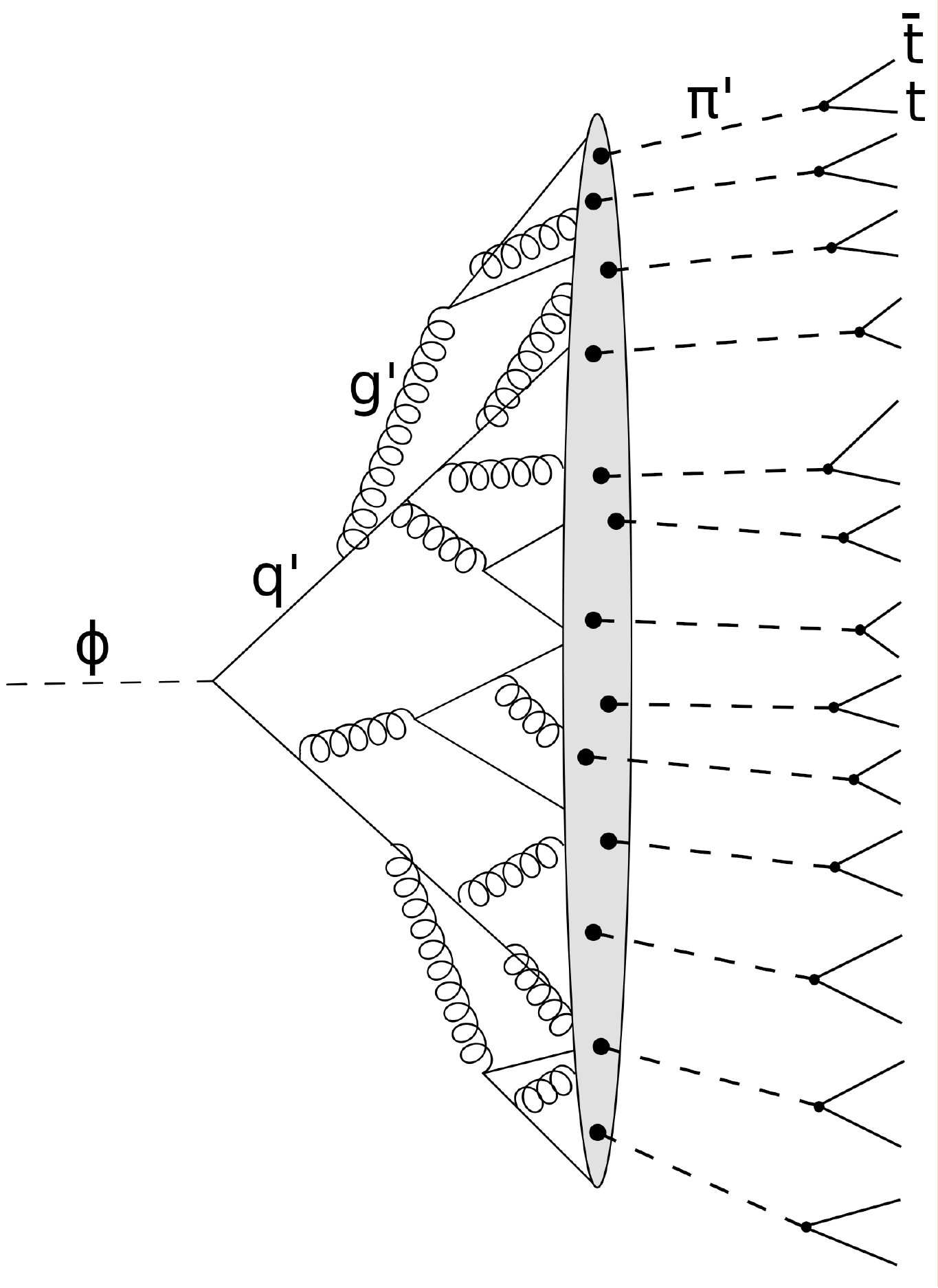}
\vspace{-0.3cm}
\caption{Our model, based on the decay of the heavy scalar $\phi$ into a shower of dark quarks and dark gluons that subsequently form dark hadrons $\pi^\prime$. The $\pi^\prime$ further decay into top quarks through portal couplings to the Standard Model.}
\label{fig:darkshower}
\vspace{-0.35cm}
\end{figure}  
\indent
Intriguingly, a recent study discovered a new contribution to $^3\overline{\text{He}}$ production through intermediate $\bar{\Lambda}_b$ resonances that generically appear in DM annihilation~\cite{Winkler:2020ltd}. This can boost the DM-induced antihelium flux enough to potentially explain AMS-02 data. This scenario is currently being investigated by several accelerator experiments~\cite{Puccio:2022}.\footnote{The key uncertainty in this scenario is the branching ratio $\bar{\Lambda}_b\rightarrow \overline{\text{He}}$ which can be measured in $pp$ collisions at the Large Hadron Collider (LHC).} However, the $\bar{\Lambda}_b$ cannot kinematically decay into $^4\overline{\text{He}}$. If some of the AMS-02 events are confirmed to be $^4\overline{\text{He}}$, a different mechanism is needed. In general, the observation of $^4\overline{\text{He}}$ is much harder to explain because standard models predict a production ratio ${^4\overline{\text{He}}}/{^3\overline{\text{He}}}\lesssim 1/1000$. One exotic production mechanism involves antimatter clouds or antimatter stars~\cite{Poulin:2018wzu}. However, the needed segregation of matter and antimatter in the galaxy is difficult to embed into a consistent cosmological model.

In this paper, we point out that an entire class of beyond Standard Model (SM) theories produces dramatically enhanced antinuclei fluxes.
Specifically, we consider SM extensions with a strongly coupled gauge sector, for instance a (heavier) version of Quantum Chromodynamics (QCD). Such models are popular because they induce high-multiplicity states of soft hadrons or leptons that escape detection at particle accelerators -- explaining the absence of new physics at the LHC (see e.g.~\cite{Strassler:2006im,Cohen:2015toa,Knapen:2016hky,Cohen:2017pzm,Pierce:2017taw,Knapen:2021eip,Albouy:2022cin}). Furthermore, the presence of additional strongly coupled gauge sectors is motivated by ultraviolet theories including superstring theory (see e.g.~\cite{Cvetic:2002qa,Cvetic:2004ui,Arkani-Hamed:2005zuc,Gmeiner:2005vz,Lebedev:2006tr,Acharya:2007rc}) and twin Higgs models~\cite{Chacko:2005pe}. 

Our main observation is that DM annihilation or decay into fermions in a strongly coupled sector produces a ``dark parton'' shower that generates a high multiplicity of the lightest strongly-coupled bound state. The subsequent decay of this particle into SM quarks via portal interactions~\cite{Galison:1983pa,Holdom:1985ag,Patt:2006fw} 
produces hundreds of soft quarks which each trigger a QCD shower. The high-multiplicity of the resulting hadrons efficiently generate antinuclei. In particular, we show that the $\overline{\text{He}}/\bar{\text{p}}$ ratio in DM annihilation/ decay is enhanced by several orders of magnitude in the presence of a strongly coupled dark sector, explaining the antihelium signal at AMS-02. More strikingly, an observable $^4\overline{\text{He}}$ flux can easily be produced.

\section{Coalescence Model}
Calculating the production of complex anti-nuclei in particle collisions is both theoretically and computationally difficult. In order to derive the spectra of antinuclei in DM annihilation (or decay), we employ a coalescence model~\cite{Schwarzschild:1963zz} in which antinucleons bind if they are produced in close proximity in phase space. The analytic coalescence model (see e.g.~\cite{Chardonnet:1997dv})
approximates the multi-antinucleon spectra as the product of single-antinucleon spectra (ignoring correlations in the antinucleon production). The differential antinucleus multiplicity $d^3N_A/ d p_A^3$ in a scattering reaction is given by: 
\be\label{eq:analytic}
E_A \frac{d^3 N_A}{dp_A^3} = 
B_A \left(E_{\bar{\text{p}}}\frac{d^3 N_{\bar{\text{p}}}}{dp_{\bar{\text{p}}}^3}\right)^{Z}\,\left(E_{\bar{\text{n}}}\frac{d^3 N_{\bar{\text{n}}}}{dp_{\bar{\text{n}}}^3}\right)^{A-Z}
\;,
\ee
evaluated at $p_{\bar{\text{p}},\bar{\text{n}}}=p_A/A$. Here, $p_A$ denotes the three-momentum of the an antinucleus with mass $A$ and charge $Z$. The coalescence factor $B_A$ accounts for the phase space volume in which antinucleons coalesce. A common definition identifies the coalescence volume with an $(A-1)$-sphere with diameter $p_c$, where $p_c$ is called the coalescence momentum,
\be\label{eq:BA}
B_A= \frac{m_A}{m_p^Z\, m_n^{A-Z}} \left( \frac{4\pi}{3} \left(\frac{p_c}{2}\right)^3\right)^{A-1}\,.
\ee
The analytic coalescence model fails in cases where correlations in antinucleon production play a role. Such correlations can be taken into account in an event-by-event coalescence model, where physical processes are simulated with a Monte Carlo generator and the coalescence condition is applied on the antinucleons of each individual event~\cite{Kadastik:2009ts}. The following event-by-event coalescence condition reproduces the predictions of the analytic coalescence model in the limit of negligible antinucleon correlations: $A$ antinucleons form a bound state if all antinucleon three-momenta evaluated in their common center-of-mass frame satisfy:\footnote{An alternative coalescence condition is to require that $A-1$ antinucleons satisfy $|p_i|<p_c/2$ in the center of mass frame of all $A$ antinucleons, while the momentum of the last antinucleon is not constrained (other than by momentum conservation). One can show that this condition leads to the same coalescence volume as Eq.~\eqref{eq:coalescence_condition} imposed on all antinucleons.}

\begin{equation}\label{eq:coalescence_condition}
 |p_i| < (A-1)^{1/(3A-3)}\, \frac{p_c}{2}\,.
\end{equation}
For the case of antideuterons ($A=2$), this simplifies to the familiar condition $|p_{1,2}|<p_c/2$ or equivalently $|p_1-p_2|<p_c$.\footnote{This condition is again applied in the two-antinucleon center-of-mass frame.} However, in the case of antihelium a non-trivial factor of $2^{1/6}$ ($^3\overline{\text{He}}$) and $3^{1/9}$ ($^4\overline{\text{He}}$) occurs in the coalescence condition. The presence of this factor (for the case of $^3\overline{\text{He}}$) was noted in~\cite{Winkler:2020ltd}, but was missed in some previous literature. We will show that Eq.~\eqref{eq:coalescence_condition} is indeed the correct event-by-event coalescence condition in App.~\ref{sec:coalescencecondition}. 

In addition to Eq.~\eqref{eq:coalescence_condition}, we require all antinucleons to stem from the same interaction vertex by imposing a cut of \mbox{$d_{\text{max}}=2\:\text{fm}$} on their relative distance. This is due to the fact that antinucleons separated by more than the nuclear radius cannot merge into an antinucleus. We note, however, that the final antinucleus fluxes are insensitive to changes in $d_{\text{max}}$ by a factor of a few.

Let us now turn to the coalescence momentum, which must be carefully chosen such that experimental data on antinucleus production -- for instance $\bar{\text{d}}$-formation in $Z$-decays at LEP~\cite{ALEPH:2006qoi} or $^3\overline{\text{He}}$-formation in $pp$-collisions at ALICE~\cite{ALICE:2017xrp,ALICE:2021mfm} -- are correctly reproduced within the coalescence model. By modulating the coalescence momentum such that all collider data is reasonably produced, the coalescence model can become relatively robust despite its myriad theoretical uncertainties. Unfortunately, due to differences in the predicted antinuclei spectra/ correlations from different Monte Carlo event generators, the determined value of $p_c$ is often a model-dependent value (see e.g.~\cite{Gomez-Coral:2018yuk}).  For Pythia 8.3~\cite{Bierlich:2022pfr}, which we employ in this work, the following coalescence momenta have been obtained for $\bar{\text{d}}$ and $^3\overline{\text{He}}$-formation~\cite{Winkler:2020ltd},
\begin{align}\label{eq:pc1}
    p_c (\bar{\text{d}}) &= 215\:\text{MeV}\nonumber\,,\\
    p_c (^3\overline{\text{He}}) &= 239\:\text{MeV}\,.
\end{align}
In the absence of experimental data on $^4\overline{\text{He}}$, we employ the following quantum mechanical argument to determine $p_c (^4\overline{\text{He}})$. Because the coalescence probability corresponds to the overlap between the multi-antinucleon product wave function and the antinucleus wave function, we obtain \mbox{$p_c(A) \propto r_A^{-1}$} with the nuclear radius $r_A$.\footnote{This approximation neglects the finite size of the hadronic interaction zone from which antinuclei are emitted.} We gain confidence in this scaling relation by observing that $p_c (^3\overline{\text{He}})/p_c (\bar{\text{d}})=1.09$ from Eq.~\eqref{eq:pc1} is in reasonable agreement with $r_{\bar{\text{d}}}/ r_{^3\overline{\text{He}}}=1.11$. Hence, we assume the following coalescence momentum for $^4\overline{\text{He}}$:
\begin{equation}
  p_c (^4\overline{\text{He}})  = \frac{r_{^3\overline{\text{He}}}}{r_{^4\overline{\text{He}}}} \:p_c (^3\overline{\text{He}}) = 281\:\text{MeV}\,.
\end{equation}

Let us remark that the coalescence model clearly has its limitations. In this work we are considering DM annihilation into final states with an enhanced (anti)nucleon density. The coalescence model omits particle-antiparticle annihilation reactions at the source which might play a role in such a case. Furthermore, there are still sizeable uncertainties in the coalescence momentum which become amplified in the predicted antihelium fluxes since e.g.\ $N_{^4\overline{\text{He}}}\propto p_c^9$. In the absence of a better description, we will nevertheless employ the coalescence model, while emphasizing that our flux predictions should not be regarded as precise. This does, however, not affect our main point that observable $^3\overline{\text{He}}$ and $^4\overline{\text{He}}$ fluxes can be obtained in a strongly coupled dark sector model.

\section{Multi-Quark Final States Induced by Dark Matter}
Our basic idea for producing a significant flux of cosmic-ray antihelium is to consider DM annihilation (or decay) into multi-quark final states. Neglecting kinematics, Eq.~\eqref{eq:analytic} suggests that the number of produced antinuclei $N_{A}$ per DM annihilation should roughly scale as
$N_{A} \propto (N_{\bar{\text{p}}})^A$,
where $A$ denotes the mass number of the antinucleus. In other words, if we increase the number of $\bar{\text{p}}$ per DM annihilation by a factor of ten, this leads to an increase of $\bar{\text{d}}$, $^3\overline{\text{He}}$, and $^4\overline{\text{He}}$ production by factors of 100, 1000 and 10000 respectively. Even if this estimate is only qualitative, the above considerations suggest that DM annihilation (or decay) into final states with many quarks will strongly enhance the cosmic-ray antinuclei fluxes relative to the antiproton flux. In the following, we present a simple and natural method to produce multi-quark final states that are capable of producing observable amounts of $^3\overline{\text{He}}$ and even $^4\overline{\text{He}}$ without violating existing antiproton constraints~\cite{Reinert:2017aga,Cuoco:2019kuu, Heisig:2020nse, DiMauro:2021qcf, Kahlhoefer:2021sha,Calore:2022stf}

Before we describe the actual model for antihelium production, it is convenient to study an analogy, which is the hadronic annihilation of WIMP DM. Let us, for the moment, assume that DM annihilates into Higgs bosons. The Higgs bosons mostly decay to bottom-antibottom quark pairs which further decay and hadronize. The resulting spectrum consists mostly of charged and neutral pions. An estimate obtained with the Pythia Monte Carlo suggests that $\mathcal{O}(100)$ pions are produced per DM annihilation. The pions are rather soft. In the rest frame of the decaying Higgs boson, their energy spectrum peaks at $E_{\pi}\sim \text{GeV}$. Including the different steps in the annihilation process we thus have,

\begin{equation}\label{eq:analogy}
\chi\chi\rightarrow hh\rightarrow 2\bar{b}b \rightarrow \mathcal{O}(100)\:\pi \,.
\end{equation}
Subsequently the pions decay to photons or light leptons. 

This example illustrates that high-multiplicity final states occur generically in the presence of a strongly coupled gauge sector. In the SM, hadronic showers are responsible for producing multi-pion final states. In beyond SM theories, it is possible to generate states of multiple heavy quarks.

For this purpose, let us turn to the model under investigation. We extend the SM by a strongly coupled dark sector which could, for instance, be a mirror version of QCD. However, the specific dark sector gauge group and field content are not important. The corresponding Lagrangian reads:
\begin{equation}
 \mathcal{L} \supset - \frac{1}{2}\,\text{Tr}\,G_{\mu\nu}^\prime G^{\prime\mu\nu}- \bar{q}^\prime (i\cancel D - m_{q^\prime}) q^\prime\,,
\end{equation}
where $G_{\mu\nu}^\prime$ is the dark gluon field strength. The dark sector contains one or several dark quark states $q^\prime$. The dark quarks and gluons bind into dark hadrons (dark mesons, dark baryons, dark glueballs etc.) at energies below a dark confinement scale $\Lambda_d$. Such a setup carries profound motivation from superstring theory and many other popular ultraviolet extensions of the SM in which the presence of additional gauge sectors is ubiquitous. Setups of this type have also been extensively studied in the context of collider physics (see e.g.~\cite{Strassler:2006im,Cohen:2015toa,Knapen:2016hky,Cohen:2017pzm,Pierce:2017taw,Knapen:2021eip,Albouy:2022cin}). 

Turning back to the analogy with WIMP annihilation, we now consider DM annihilation (or DM decay) into pairs of heavy scalars $\phi$ that could, e.g.,\ be the heavy Higgs bosons of an extended Higgs sector. Instead of directly decaying into SM particles, the heavy scalars decay into pairs of dark quarks (via a Yukawa coupling $\mathcal{L}\supset y_\phi \phi \bar{q}^\prime q$). These induce a shower of dark partons that subsequently bind into dark hadrons. Heavy dark hadrons further cascade down to lighter dark hadron states which we call $\pi^\prime$.\footnote{The $\pi^\prime$ could be dark mesons, dark glueballs or onium states depending on the specific model.} This process which is illustrated in Fig~\ref{fig:darkshower} is very similar to the production of pions in a QCD shower (cf.\ Eq.~\eqref{eq:analogy}). The analogy with QCD also suggests that we can expect 
\begin{equation}
N_{\pi^\prime} = \mathcal{O}(100)    
\end{equation}
for the number of $\pi^\prime$ generated in the dark parton shower. However, due to the sensitivity to the particle content and (running) gauge coupling of the hidden sector, the number $N_{\pi^\prime}$ could also easily be an order of magnitude different from $N_\pi$ in the QCD example (see e.g.~\cite{Knapen:2016hky}). For instance, quasi-conformal strongly-coupled dark sectors are known to yield higher $N_{\pi^\prime}$ compared to QCD-like models~\cite{Hatta:2008tn}. Furthermore, $N_{\pi^\prime}$ is very sensitive to the mass of $\pi^\prime$ in relation to the dark confinement scale $\Lambda_d$. 

A key difference between the $\pi^\prime$ and SM pions is that the $\pi^\prime$ can be much heavier if $\Lambda_d \gg \Lambda_{\text{QCD}}$. This is a plausible assumption, because it explains why the hidden sector has so far escaped detection at particle accelerators. For example, we may assume a DM mass $m_\chi \sim m_\phi \sim 1000\:\text{TeV}$ and $\Lambda_d\gtrsim m_{\pi^\prime}\sim \text{TeV}$.  

In the final step, the $\pi^\prime$ decays into visible matter through one of the portals of the SM, for instance the vector portal or the Higgs portal~\cite{Galison:1983pa,Holdom:1985ag,Patt:2006fw}. The antinuclei production is not particularly sensitive to the specific decay mode of $\pi^\prime$ as long as it is hadronic. This is due to the fact that any type of quark or gluon final state induces a QCD shower leading to a comparable antinucleon spectrum. For concreteness we will assume a two-body decay $\pi^\prime \rightarrow \bar{t}t$ (see e.g.~\cite{Strassler:2008fv} for a model in which dark pions decay into pairs of heavy-flavor quarks).\footnote{Couplings of $\pi^\prime$ to top-quarks are also experimentally less constrained compared to couplings to light flavor states.} The full DM annihilation process we consider is thus (see Fig.~\ref{fig:darkshower})
\begin{equation}\label{eq:process}
\chi\chi \rightarrow \phi\phi\rightarrow 2\bar{q}^\prime q^\prime \rightarrow N_{\pi^\prime}\; \pi^\prime \rightarrow N_{\pi^\prime} \;\bar{t}t\,,
\end{equation}
where we will set $N_{\pi^\prime}=100-1000$ as expected for a strongly coupled hidden sector gauge group. Similarly, we will also consider decaying DM for which one simply has to replace $\chi\chi$ by $\chi$ in the above process. 

Each of the many top quarks produces a shower of hadrons -- mostly pions, but also a considerable number of antinucleons. In the final step, antiprotons and antineutrons with low relative momentum bind into antideuterons and antihelium nuclei. As we argued at the beginning of this section, the high-antinucleon-multiplicity states we expect in this model can trigger a dramatic enhancement of the antihelium flux from DM annihilation (relative to the antiproton flux).

We note one final important consideration: the antinucleons induced by DM can only merge into an antinucleus if their physical distance at production is comparable to or smaller than the nuclear radius, i.e.\ if $d\lesssim d_{\text{max}} =2\:\text{fm}$. In the considered model, the $\pi^\prime$ are produced promptly since $\Lambda_d \gg \Lambda_{\text{QCD}}$.\footnote{The typical size of the QCD' vertex is given by $\Lambda_d^{-1} \ll \text{fm}$.} However, the small-distance condition additionally requires a sufficiently prompt $\pi^\prime$-decay, more specifically a decay width $\Gamma_{\pi^\prime}\gtrsim \text{GeV}$. This suggests a relatively heavy $\pi^\prime$ around/ above the TeV-scale for which such a large decay width can be achieved without unreasonably large portal couplings~\cite{Knapen:2021eip}.

\section{Model Implementation}\label{sec:implementation}

The full implementation of particle physics models with a strongly coupled hidden sector -- including a dark showering algorithm -- unfortunately goes far beyond the scope of this paper.\footnote{While existing Pythia models include gauge extension of the SM, these are distinct from the class of models we investigate here.} Instead, we have performed a ``toy implementation'' which aims at capturing the main features of the process~\eqref{eq:process}. 

Specifically, we added $\phi$, $\pi^\prime$, and $n$ auxiliary resonances $\varphi_i$ (which stand for some intermediate partons/hadrons in the dark shower) to the field content of the SM. Then we built an approximate version of a dark parton shower by implementing the following decay chain into Pythia,
\begin{equation}
\phi\rightarrow 2\varphi_{1} \rightarrow 4\varphi_{2} \rightarrow \dots 2^n \varphi_n \rightarrow 2^{n+1} \pi^\prime \rightarrow 2^{n+1} \bar{t}t\,.
\end{equation}
In this way we can simulate the production of a large number of $\pi^\prime$ which subsequently decay into $\bar{t}t$. In order to realize multiplicities of $N_{\pi^\prime}\simeq 100-1000$, we include $n=6-9$ of the auxiliary resonances $\varphi_i$ in the shower.

For the masses of each $\varphi_i$, we follow the simple pattern $m_{\phi}=r\, m_{\varphi_1} = \dots = r^n m_{\varphi_n}=r^{n+1}\,m_{\pi^\prime}$. Because we expect the $\pi^\prime$ from the dark parton shower to be relatively soft ($E_{\pi^\prime}\lesssim 10 m_{\pi^\prime}$ in the rest frame of $\phi$), we utilize values for $r\simeq 2-2.5$. The resulting antihelium fluxes change by $1-2$ orders of magnitude if we vary $r$ within this range, where the maximum fluxes are obtained for small $r$. However, this variation is small compared to the overall enhancement of the antihelium fluxes and can easily be compensated by a slight change of the multiplicity $N_{\pi^\prime}$ (a harder $\pi^\prime$-spectrum yields the same antinucleus production for a slightly higher $N_{\pi^\prime}$). For concreteness we choose $r=2.01$ in our default implementation, a result that maximizes the antinuclei production. 

We note one conservative difference between our toy-model and a true dark parton shower. Our toy model generates an approximately isotropic distribution of $\pi^\prime$ (in the rest frame of $\phi$), while a QCD-like dark sector would induce dark hadrons within two jet-like structures. Since antinuclei are formed more efficiently in jets~\cite{Kadastik:2009ts}, we expect that our toy model would underestimate the antinuclei production-- which compensates for the somewhat optimistic choice of $r$.

The $\bar{t}t$-pairs resulting from $\pi^\prime$-decay further decay and fragment into hadrons including antinucleons. This part of the decay chain is treated by Pythia's standard hadronization algorithm. In the final step we derive the antinucleus production by applying the coalescence condition~\eqref{eq:coalescence_condition} on the produced antinucleons on an event-by-event basis. In the case of $^3\overline{\text{He}}$ we also include the contribution from antitriton decay. 

While our Pythia implementation is designed to model antinucleus formation in dark parton showers, it also yields correct predictions for models with dark matter annihilation into multi-step cascades of mediators~\cite{Elor:2015tva} with a mass scale set such that each mediator has approximately half the mass of its heavier partner. However, since such cascade models are tuned, the production of high-multiplicity antinucleon states in dark parton showers appears more attractive.

Let us remark that we only treat the decay of $\phi$ in Pythia. In order to obtain the antinucleus spectrum from DM annihilation we can simply multiply the spectrum from $\phi$-decay by a factor of two and apply a Lorentz boost with a Lorentz factor of $\gamma = m_\chi/m_\phi$ (or $\gamma = 0.5\,m_\chi/m_\phi$ in the case of decaying DM). This is possible since $\phi$ itself does not carry color or dark color charge. Since the antinucleus spectrum in the $\phi$ rest frame is typically very soft, we find that the shape of the boosted spectrum is mostly determined by $m_\chi/m_\phi$. This makes our results more robust since any errors in the antinuclei spectral shape -- which may be caused by our simplistic implementation of the dark parton shower -- are washed out by the Lorentz boost and do not propagate to the final fluxes. Nevertheless, we should emphasize that normalization of the predicted antihelium fluxes should only be seen as an estimate. However, since large uncertainties anyway exist within the coalescence description, this is sufficient for our purposes.

\section{Calculation of Antinuclei Fluxes}
We consider both DM annihilation and DM decay via the process given in Eq.~\eqref{eq:process} (with $\chi\chi$ replaced by $\chi$ in the case of decay). The antinucleus energy spectrum per annihilation or decay event $dN_{A}/dE_A$ is obtained from Pythia as described in Sec.~\ref{sec:implementation}. The source term in the galactic halo reads:
\begin{align}
Q&= \frac{1}{2} \left(\frac{\rho_\chi}{m_\chi}\right)^2 \langle\sigma v \rangle\; \frac{dN_A}{dE_A}\quad &\text{(annihilation)}\,,\nonumber\\
Q&= \left(\frac{\rho_\chi}{m_\chi}\right) \Gamma\; \frac{dN_A}{dE_A}\quad &\text{(decay)}\,,
\end{align}
where $\rho_\chi$ is the DM density, for which we assume a Navarro–Frenk–White profile~\cite{Navarro:1995iw} with a local density of $\rho_0=0.38\:\text{GeV}/\text{cm}^3$~\cite{McMillan:2016}. For annihilating DM, the source term scales with the velocity-averaged annihilation cross section $\langle \sigma v\rangle$, while decaying DM scales with the decay rate, $\Gamma$.

To obtain the local antinuclei fluxes, we propagate the cosmic rays through the Milky Way. Since propagation is not the focus of our study, we use a standard two-zone diffusion model fit to AMS-02 B/C and $\bar{\text{p}}$ data~\cite{AMS:2021nhj} as described in~\cite{Heisig:2020nse,DiMauro:2021qcf} (specifically we choose propagation parameters from Table~V in~\cite{DiMauro:2021qcf} adjusted to a diffusion halo of $L=10\:\text{kpc}$). For propagation in the heliosphere we apply the force-field approximation~\cite{Gleeson:1968zza} with a Fisk potential of $\phi=600\:\text{MV}$ -- which is typical for AMS-02 data (see e.g.~\cite{Corti:2015bqi}).

\section{Results}
We now present results for the DM-induced antinuclei fluxes for two versions of the strongly-coupled dark sector model: one assuming DM annihilation through the process in Eq.~\eqref{eq:process} and the other assuming DM decay through the same process with $\chi\chi$ replaced by $\chi$. The choice of the model parameters -- the masses $m_\chi$, $m_\phi$, $m_{\pi^\prime}$, the multiplicity $N_{\pi^\prime}$, and the annihilation cross section $\langle\sigma v\rangle$ or decay rate $\Gamma$ -- are listed in Tab.~\ref{tab:antinucleus}. Note that we are considering very heavy DM particles with $m_\chi = 150\:\text{TeV}$ or $m_\chi = 5000\:\text{TeV}$. This is required by kinematics since the dark parton showers induce final states with $100-1000$ $\pi^\prime$ that decay into tops with mass $m_t=172\:\text{GeV}$. Because the number density of DM scales as $n_\chi \propto m_\chi^{-1}$, these heavy masses suppress standard indirect searches for $\gamma$-rays and antiprotons (e.g.~\cite{Montanari:2022buj}).

\begin{table}[tp]
\begin{tabular}{|ccc|}
\hline && \\[-2.5mm]
DM type & Annihilating  & Decaying \\[1mm] \hline & & \\[-3mm]
\multicolumn{3}{|c|}{Input Parameters}\\[0.5mm] \hline & & \\[-3mm]
$m_\chi$~[TeV] & 150 & 5000\\[1mm]
$m_\phi$~[TeV] & 50.4& 375\\[1mm]
$m_{\pi^\prime}$~[GeV] & 380 & 700\\[1mm]
$N_{\pi^\prime}$ & $256$ & $1024$\\[1mm]
$\quad\langle \sigma v \rangle$~[cm$^3$s$^{-1}$]$\quad$ & $\quad 6.6\times 10^{-24}\quad$& $-$\\[1mm]
$\Gamma$~[s$^{-1}$]& $-$ & $\quad 9\times 10^{-30}\quad$\\[1mm]
 \hline  & & \\[-3mm]
\multicolumn{3}{|c|}{Antinuclei Events at AMS-02}\\[0.5mm] \hline & & \\[-2.5mm]
$^3\overline{\text{He}}$ & $15.6$ & $20.3$\\[1.5mm]
$^4\overline{\text{He}}$ & $1.0$ & $3.1$\\[1.5mm]
$\bar{\text{d}}$ & $19.3$ & $1.2$ \\[1mm] \hline  & & \\[-3mm]
\multicolumn{3}{|c|}{Antinuclei Events at GAPS}\\[0.5mm] \hline & & \\[-2.5mm]
$\bar{\text{d}}$ & $0.7$ & $0$ \\[1mm] \hline
\end{tabular}
\caption{Input parameters of one annihilating and one decaying DM benchmark scenario. Also given are the predicted antihelium and antideuteron event numbers at AMS-02 (per ten years) and GAPS.}
\label{tab:antinucleus}
\end{table}

For the example of annihilating DM we, therefore, choose a DM annihilation cross section close to the unitarity limit\footnote{We state the unitarity limit for s-wave annihilation. The general bound contains an extra factor of ($2J+1$) with $J$ denoting the angular momentum.},
\begin{equation}
(\sigma v)_{\text{unit.}} = \frac{4\pi}{m_{\chi}^2 v}\,,
\end{equation}
in order to maximize antinuclei fluxes. We note that the relative DM velocity $v$ at freeze out ($v\sim 0.3$) is very different from the galactic halo velocity, $v\sim 10^{-3}$. Thus, our annihilating DM model (with a cross section from Tab.~\ref{tab:antinucleus}) is compatible with a thermal origin of DM if we assume $\sigma v\propto 1/v$. Such a cross section scaling is commonly obtained in scenarios with Sommerfeld enhancement~\cite{Sommerfeld:1931qaf,Hisano:2004ds}.\footnote{The correct relic abundance of a $150\:\text{TeV}$ DM particle with a unitary annihilation cross section is implied by the unitarity limit on the DM mass~\cite{Griest:1989wd}.} 

While the unitarity limit imposes an upper bound on the observable cosmic-ray flux, we should emphasize that several loopholes exist (see e.g.~\cite{Tak:2022vkb}). For instance, the unitarity limit does not apply to composite DM~\cite{Harigaya:2016nlg} -- a natural possibility in strongly coupled dark sectors -- for which the maximal cross section is set by the geometric size. 

Furthermore, no such theoretical constraints apply to decaying DM, for which the unitarity limit is irrelevant. In this case, the leading constraint arises from the antiproton channel (which we include in our analysis) and requires a lifetime considerably longer than the age of the universe. Such a long lifetime can be ensured by an approximate symmetry.\footnote{For instance, the stability of $\chi$ can be protected by a classical symmetry which is broken by quantum effects. In such a case $\chi$ decays at an exponentially suppressed rate.}. As an example, we consider $m_\chi = 5000\:\text{TeV}$. While such ultra-heavy DM particles cannot have a thermal origin, many plausible non-thermal production mechanisms exist (including production during reheating, gravitational production, production in a first-order phase transition)~\cite{Kolb:1998ki}. 

\begin{figure*}[htp]
\centering
\includegraphics[width=.47\textwidth]{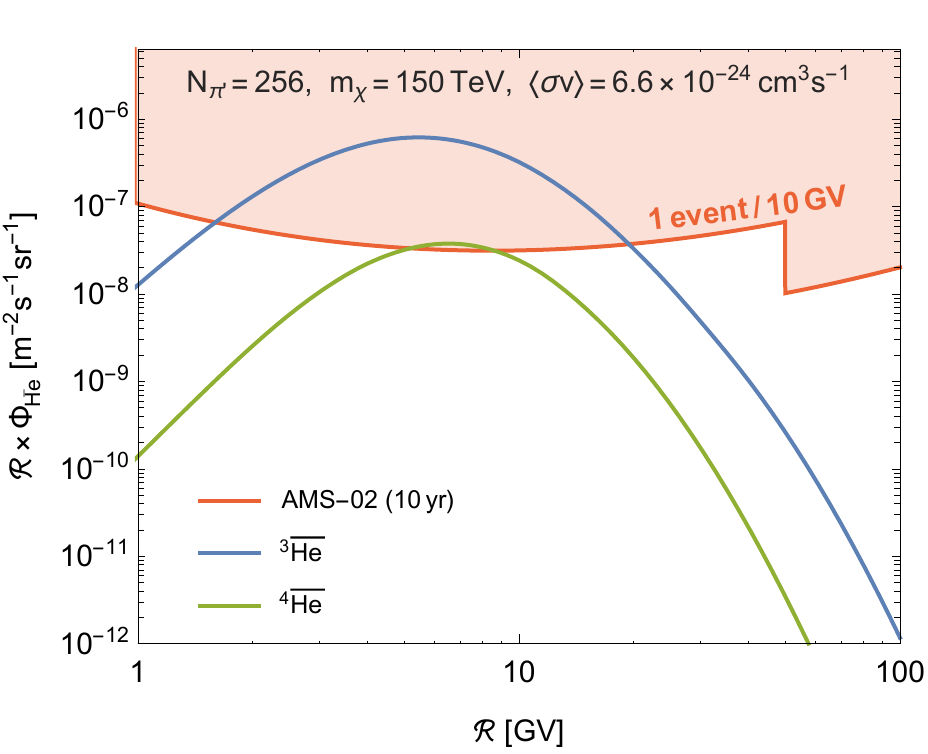}\hspace{2mm}
\includegraphics[width=.47\textwidth]{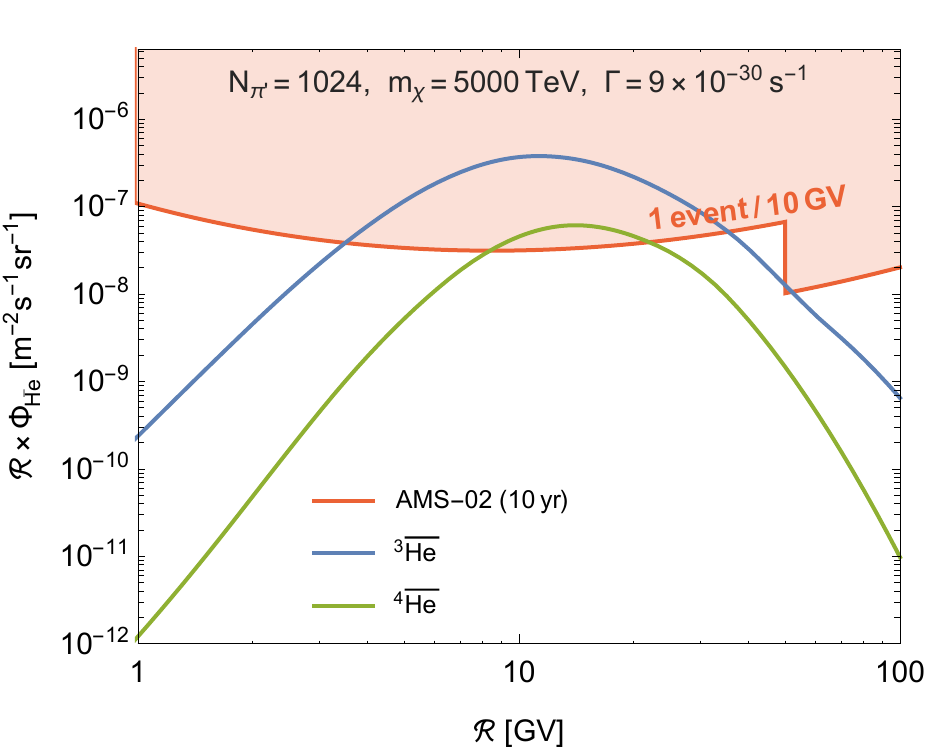}\\[1mm]
\includegraphics[width=.47\textwidth]{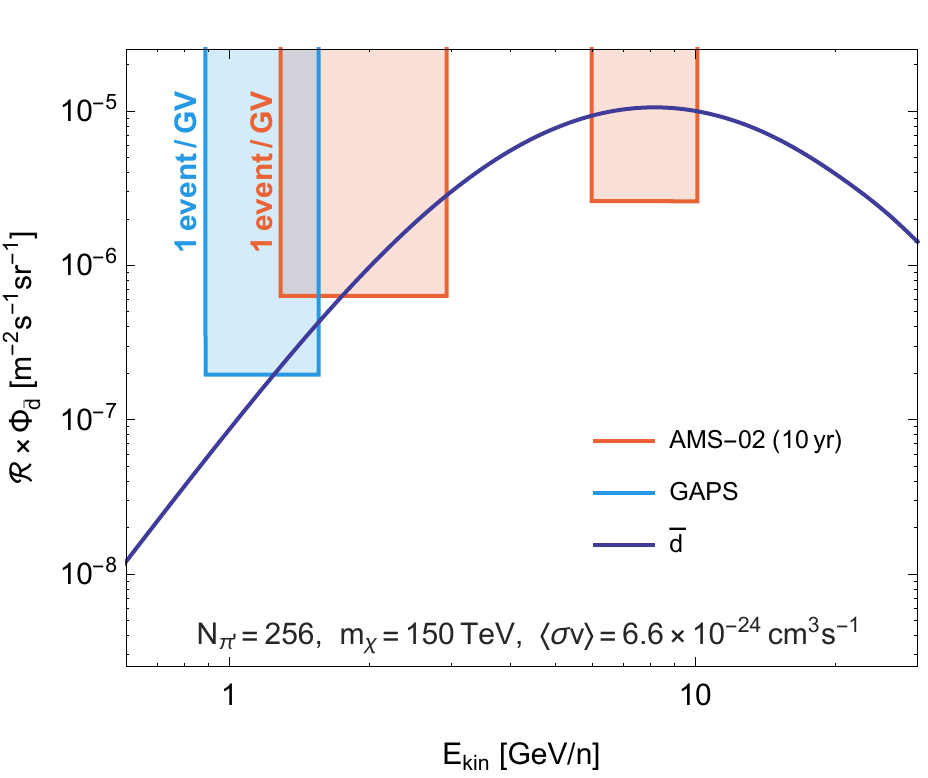}\hspace{2mm}
\includegraphics[width=.47\textwidth]{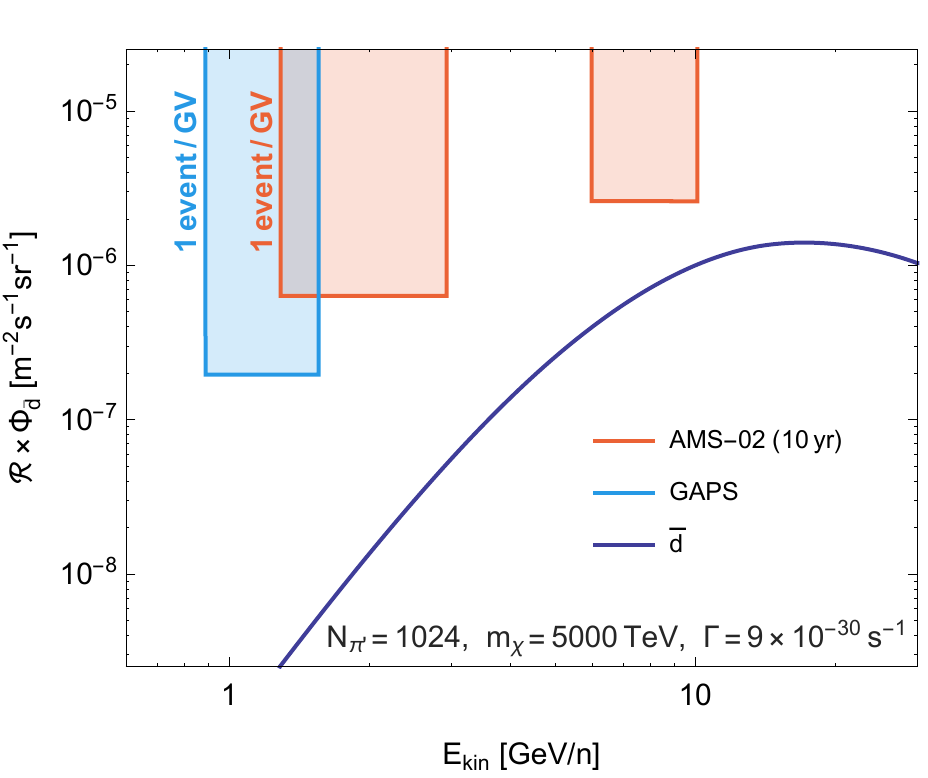}\\[1mm]
\includegraphics[width=.47\textwidth]{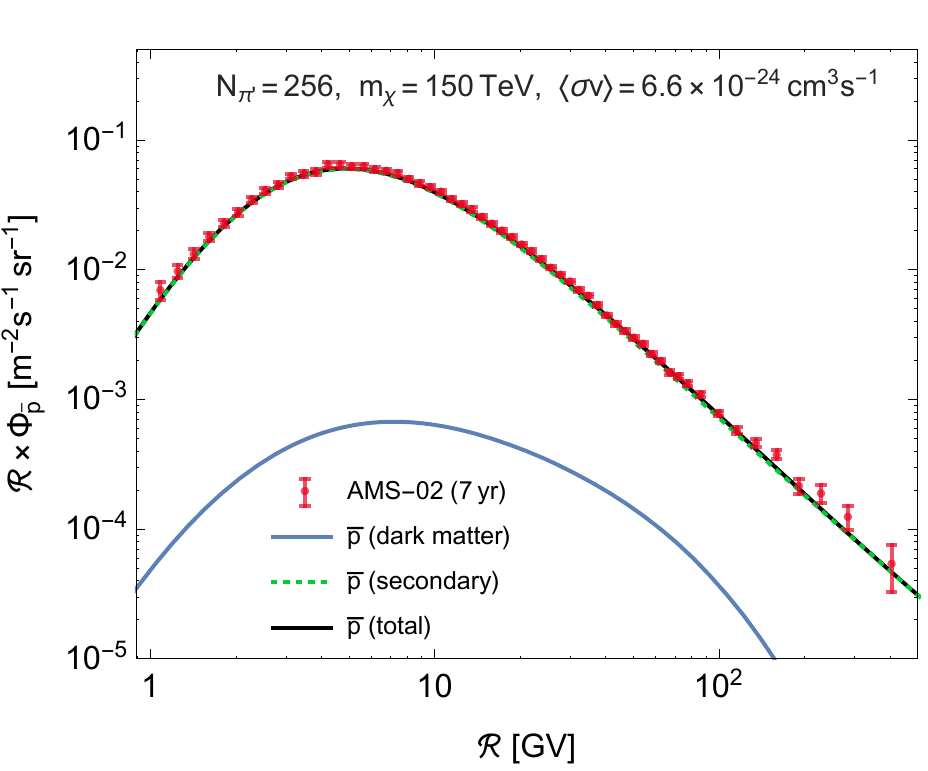}\hspace{2mm}
\includegraphics[width=.47\textwidth]{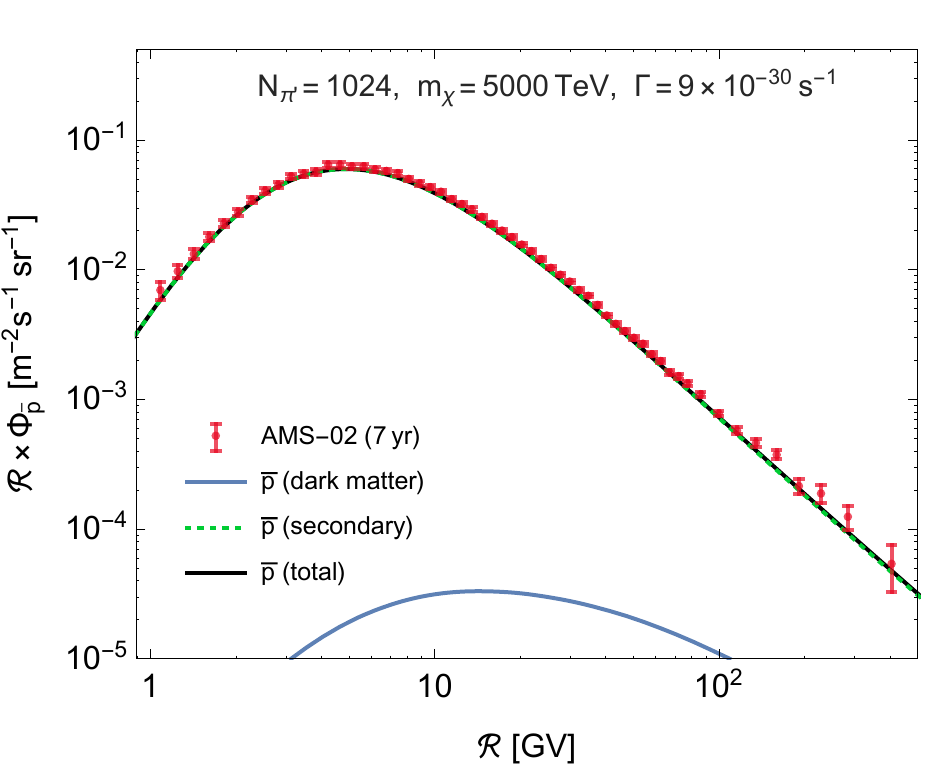}
\caption{Antihelium (upper panels), antideuteron (middle panels) and antiproton (lower panels) fluxes in dark matter models with a strongly coupled dark sector. The dark matter annihilation/ decay process is given in Eq.~\eqref{eq:process} and depicted in Fig.~\ref{fig:darkshower}. The left panels refer to the case of annihilating dark matter, the right panels to the case of decaying dark matter. Also depicted are the projected AMS-02 and GAPS antinuclei sensitivities. The predicted number of antinuclei events are given in Tab.~\ref{tab:antinucleus}. For the antiproton channel, the AMS-02 data and the predicted secondary astrophysical background are also shown.}
\label{fig:fluxes}
\end{figure*}  

Let us now compare the antinuclei production in the strongly coupled dark sector model with standard scenarios. For the annihilating DM benchmark example, our Pythia simulations indicate production ratios of 
\begin{equation}
    \bar{\text{p}}:\bar{\text{d}}:{^3\overline{\text{He}}}:{^4\overline{\text{He}}}= 3\times 10^4 : 3\times 10^2 : 18 :1\,.
\end{equation}
In comparison, the ratios achieved in astrophysical processes and in standard WIMP DM annihilation are roughly
\begin{equation}
    \bar{\text{p}}:\bar{\text{d}}:{^3\overline{\text{He}}}:{^4\overline{\text{He}}}= 10^{10} : 10^7 : 10^4 :1\,.
\end{equation}
The enhancement of $^4\overline{\text{He}}$-production relative to $\bar{\text{p}}$ thus reaches six orders of magnitude in the strongly coupled dark sector models compared to standard processes. In fact the enhancement is even a bit larger in the decaying DM example.

In Fig.~\ref{fig:fluxes} we present the cosmic ray fluxes of $\bar{\text{p}}$, $\bar{\text{d}}$, $^3\overline{\text{He}}$ and $^4\overline{\text{He}}$ obtained in the annihilating and decaying DM scenarios of Tab.~\ref{tab:antinucleus}. Furthermore, we show the AMS-02 antiproton data~\cite{AMS:2021nhj} and projected antinuclei sensitivities~\cite{Kounine:2011,Winkler:2020ltd}.\footnote{The depicted AMS-02 antinuclei sensitivities rely on pre-launch estimates provided by the AMS-02 collaboration~\cite{Giovacchini:2007dwa,Kounine:2011}. They may not fully reflect later changes in the detector configuration and analysis details.} We also include the GAPS sensitivity to antideuterons~\cite{Aramaki:2015laa}. The number of expected antinucleus events is given in Tab.~\ref{tab:antinucleus}.

Strikingly, the expected antihelium fluxes are above AMS-02 sensitivity in the annihilating and decaying DM examples. Both scenarios predict $\sim 20$ antihelium events per ten years, consistent with the tentative antihelium signal at AMS-02~\cite{Ting:2016,Choutko:2018,vonDoetinchem:2020vbj} (which features 8 events in the first $\sim 5$ years of data). A particularly thrilling observation is that even the potential detection of $^4\overline{\text{He}}$ can be accommodated in the strongly coupled dark sector model. The decaying DM benchmark model predicts a slightly larger number of $3.1$ $^4\overline{\text{He}}$ events compared to $1.0$ $^4\overline{\text{He}}$ event for the annihilating DM scenario. We note that the relative importance of $^4\overline{\text{He}}$ increases with $N_{\pi^\prime}$. The annihilating DM example was chosen to be compatible with a thermal origin of DM which imposes $N_{\pi^\prime}\lesssim \text{few}\times 100$. Otherwise $m_\chi$ would exceed the unitarity limit~\cite{Griest:1989wd}. If we drop the assumption of thermal production -- as in the decaying DM benchmark -- $N_{\pi^\prime}\sim 1000$ can easily be realized which translates to the even higher $^4\overline{\text{He}}$-flux. Finally, we note that the ratio of $^4\overline{\text{He}}$ to $^3\overline{\text{He}}$ events strongly depends on the assumed coalescence models for each antinucleus, and thus these ratios should be considered as best-estimates.

The antideuteron flux in the annihilating DM case falls into the sensitivity window of AMS-02 ($19$ predicted events per 10 years) and GAPS ($0.7$ predicted events). Since antideuteron searches are, however, limited to low rigidities $\mathcal{R}<10\:\text{GV}$, the observability of antideuterons depends on the exact kinematics. For the decaying DM case, which features a harder spectrum (due to the larger $m_\chi/m_\phi$ ratio), more of the predicted antideuteron flux is above the rigidity threshold, and the normalization is lower. Hence, we predict only a single event at AMS-02 and none at GAPS. Hence, an observable antideuteron signal may occur in the strongly coupled dark sector models, but it is less generic than the antihelium signal.

Turning finally to the antiproton channel, we observe that the flux induced by DM annihilation or decay is strongly subdominant compared to the astrophysical antiproton background which is also shown in Fig.~\ref{fig:fluxes} (taken from~\cite{DiMauro:2021qcf}). The quality of the antiproton fit is only marginally affected by the DM component. In fact, in both benchmark examples, we find a small improvement of the fit by $\Delta\chi^2\sim 2$. Due to the hadronic dark matter annihilation/ decay complementary indirect detection channels (gamma rays, positrons) are expected to be less sensitive compared to antiprotons.

\section{Discussion and Conclusions}

We have shown that the annihilation or decay of heavy DM through a strongly coupled dark sector can produce an observable flux of cosmic-ray $\bar{\text{d}}$, $^3\overline{\text{He}}$ and $^4\overline{\text{He}}$, while remaining consistent with all current indirect detection 
constraints. This result is of significant interest due the tentative observation of approximately six $^3\overline{\text{He}}$ and two $^4\overline{\text{He}}$ particles by the AMS-02 collaboration~\cite{Ting:2016,Choutko:2018,vonDoetinchem:2020vbj}.\footnote{The exact number of detected events and the isotopic distribution have not been stated officially by the AMS-02 collaboration, and different numbers have been quoted in the literature.} The production of $^4\overline{\text{He}}$, in particular, has been extremely difficult to explain with astrophysical~\cite{Shukla:2020bql}, dark matter~\cite{Cholis:2020twh, Winkler:2020ltd}, or even exotic cosmological models~\cite{Poulin:2018wzu}. Therefore, the confirmation of even a single $^4\overline{\text{He}}$ event by AMS-02 would constitute one of the most striking observations in astroparticle physics. In this article, we argued that such a discovery would most likely hint at the production of high-multiplicity antinucleon states that generically appears in dark matter models with strongly coupled gauge sectors. The presence of such gauge sectors carries profound motivation from leading beyond-SM theories. Needless to say that the prospect that cosmic ray $^4\overline{\text{He}}$ could not only reveal the nature of dark matter, but also shine light on the correct particle theory at high energy is extremely exciting. One could even imagine to employ future-observed antinuclei isotopic ratios to reveal the properties of the dark sector gauge group.

Our study challenges a long-standing lore that the observation of heavier cosmic-ray antinuclei would be an unambiguous sign of antimatter domains in our galaxy~\cite{Chardonnet:1999bq}. In fact, our strongly-coupled dark sector models could not only induce observable $\bar{\text{d}}$, $^3\overline{\text{He}}$ and $^4\overline{\text{He}}$ fluxes, but even a non-negligible $^6\overline{\text{Li}}$ flux. While the latter likely falls below the AMS-02 sensitivity - naive estimates suggest a factor of $100-1000$ suppression compared to  $^4\overline{\text{He}}$ -- it is potentially within reach of future missions like AMS-100~\cite{Schael:2019lvx} and ALADInO~\cite{Battiston:2021org}.

Collider searches offer another exciting pathway to test the high-multiplicity antinucleon states emerging from strongly-coupled dark sector models. In order to strongly enhance  antinuclei fluxes, the lightest dark sector hadrons $\pi^\prime$ should decay promptly into SM quarks. This is due to the fact that only antinucleons from the same decay vertex can merge into an antinucleus due to the limited range of the nuclear force. The necessity of prompt $\pi^\prime$-decay implies significant portal interactions between the dark and the visible sectors. 
For the considered dark matter annihilation/ decay process (Eq.~\eqref{eq:process}), these manifest in an effective $\pi^\prime\bar{t}t$-coupling which can induce gluon fusion production of $\pi^\prime$ at the LHC and subsequent decay to ditops. The resulting signature in the ditop invariant mass distribution can potentially be tested in future CMS and ATLAS searches. In fact, for $m_{\pi^\prime}\sim 400\:\text{GeV}$ and an optimistic choice of the effective $\pi^\prime\bar{t}t$-coupling, the model can accommodate an observed moderate excess in the CMS ditop-channel~\cite{CMS:2019pzc,Arganda:2021yms}.
But it would be even more exciting if the entire dark matter annihilation/decay chain could be probed in a collider experiment. For the heavy dark matter scenario studied in this work this will require a 100\:\text{TeV}-collider~\cite{Arkani-Hamed:2015vfh} (or beyond). However, variations of our models could reside at lower energy scales and induce enhanced antinuclei fluxes by $\pi^\prime$-decays to lighter quarks. In such cases one could hope to directly observe the high-multiplicity states induced by dark hadron showers with dedicated triggers at the LHC~\cite{Knapen:2016hky,Cohen:2017pzm}.

Finally, while the dramatic enhancement of cosmic ray $^3\overline{\text{He}}$ and $^4\overline{\text{He}}$ is a robust prediction of the strongly coupled gauge extensions of the Standard Model, our flux predictions should not be seen as precise. Our Monte Carlo implementation of dark hadron showers is rather simplistic, leaving room for future improvements. Furthermore, the exact antinuclei fluxes depend sensitively on the presumed coalescence model, which is poorly constrained for $^3\overline{\text{He}}$ and can only be theoretically estimated for $^4\overline{\text{He}}$. Given that an antihelium discovery in cosmic rays could be just around the corner, dedicated experimental and theoretical efforts to improve the description of antinucleus formation and to precisely pin down the predictions of strongly coupled dark sector models are urgently needed. 

\section*{Acknowledgements}
We would like to thank Ilias Cholis, Tim Cohen, Philip von Doetinchem, Pierre Salati, and Paolo Zuccon for helpful comments. MWW\ acknowledges support by the Swedish Research Council (Contract No.\ 638-2013-8993). PDTL and TL are supported by the Swedish Research Council under contract 2019-05135 and the European Research Council under grant 742104. TL is also supported by the Swedish National Space Agency under contract 117/19. This project used computing resources from the Swedish National Infrastructure for Computing (SNIC) under project Nos. 2021/3-42, 2021/6-326 and 2021-1-24 partially funded by the Swedish Research Council through grant no. 2018-05973.

\appendix

\section{Event-by-event coalescence condition for Antinuclei}\label{sec:coalescencecondition}
In this appendix we proof the event-by-event coalescence condition~\eqref{eq:coalescence_condition}. For this purpose we assume the absence of correlations in the antinucleon production spectra such that the multi-antinucleon spectra are given by product of single-antinucleon spectra. In this limit the predictions of the event-by-event coalescence model and the analytic coalescence model have to converge. We can write the differential antinucleus multiplicity as (see e.g.~\cite{Fornengo:2013osa})
\begin{align}\label{eq:coalescenceN}
 \frac{d^3 N_{A}}{dp_A^3} =&  \left(\frac{d^{3} N_{\bar{\text{p}}}}{d p_{\bar{\text{p}}}^3}\right)^Z \left(\frac{d^{3} N_{\bar{\text{n}}}}{d p_{\bar{\text{n}}}^3}\right)^{A-Z}\times \nonumber\\
 &\int \left(\prod\limits_i d^3 p_i\right)  \mathcal{P}(p_i) \; \delta^{(3)}\!\left(-p_A+\sum\limits_i p_i\right)
\end{align}
where $\mathcal{P}(p_i)$ denotes the probability that a system of $A$ antinucleons with three-momenta $p_i$ forms an antinucleus. We employed that $\mathcal{P}(p_i)$ is non-zero only at $p_i \simeq p_A/A\equiv p_{\bar{\text{p}},\bar{\text{n}}}$ such that $d^3N_{\bar{\text{p}},\bar{\text{n}}}/dp_i^3 \simeq d^3N_{\bar{\text{p}},\bar{\text{n}}}/dp_{\bar{\text{p}},\bar{\text{n}}}^3$. This replacement allowed us to pull the differential antinucleon multiplicities out of the momentum integral. 

It is convenient to rewrite Eq.~\eqref{eq:coalescenceN} as 
\begin{align}\label{eq:coalescenceN2}
 E_A\frac{d^3 N_{A}}{dp_A^3} =&  \left(E_{\bar{\text{p}}}\frac{d^{3} N_{\bar{\text{p}}}}{d p_{\bar{\text{p}}}^3}\right)^Z \left(E_{\bar{\text{n}}}\frac{d^{3} N_{\bar{\text{n}}}}{d p_{\bar{\text{n}}}^3}\right)^{A-Z}\times \mathcal{I}_A\,,
\end{align}
with
\begin{align}
   \mathcal{I}_A = \frac{E_A}{E_{\bar{\text{p}}}^{Z} E_{\bar{\text{n}}}^{A-Z}}\int & \left(\prod\limits_i d^3 p_i\right)  \times\nonumber\\
   &\mathcal{P}(p_i) \; \delta^{(3)}\!\left(-p_A+\sum\limits_i p_i\right)\,.
\end{align}  
The integral $\mathcal{I}_A$ is Lorentz-invariant. We can conveniently evaluate it in the joined antinucleon center-of-mass frame (defined by $\sum\limits_i p_i=0$) and the result remains valid in any frame. Plugging in the 
coalescence condition~\eqref{eq:coalescence_condition}, we obtain
\begin{align}
 \mathcal{I}_A = &\frac{m_A}{m_{\bar{\text{p}}}^{Z} m_{\bar{\text{n}}}^{A-Z}}\:\times\nonumber\\
  &\int  \left(\prod\limits_i d^3 p_i \: \Theta\left( p_c^*- |p_i| \right)\right) \delta^{(3)}\!\left(\sum\limits_i p_i\right)\,,
\end{align}
where we employed that all momenta are non-relativistic in the regime, where the coalescence condition is fulfilled (in this frame). Furthermore, we introduced
\begin{equation}
 p_c^*= (A-1)^{1/(3A-3)}\: p_c\,.
\end{equation}
Evaluating the integral yields
\begin{align}\label{eq:IAfinal}
    \mathcal{I}_A &\simeq \frac{m_A}{m_{\bar{\text{p}}}^{Z} m_{\bar{\text{n}}}^{A-Z}}\frac{1}{A-1} \left[ \frac{4\pi}{3} \left(\frac{p_c^*}{2}\right)^3\right]^{A-1}\nonumber\\
    &= \frac{m_A}{m_{\bar{\text{p}}}^{Z} m_{\bar{\text{n}}}^{A-Z}}\left[ \frac{4\pi}{3} \left(\frac{p_c}{2}\right)^3\right]^{A-1}=B_A\,,
\end{align}
where we employed the definition of $B_A$ from Eq.~\eqref{eq:BA} in the last step. Plugging Eq.~\eqref{eq:IAfinal} into Eq.~\eqref{eq:coalescenceN2} reproduces the formula for antinucleus production in the analytic coalescence model (Eq.~\eqref{eq:analytic}). Hence, we conclude that Eq.~\eqref{eq:coalescence_condition} is the correct event-by-event coalescence condition.

\bibliography{main}

\end{document}